\documentclass[11pt]{article}
\usepackage[utf8]{inputenc}
\usepackage[T2A]{fontenc}               
\usepackage[russian,english]{babel}                 
\usepackage{epsfig,graphics,color}
\author{ \\
\large \bf  Zafar U. Usubov\footnote      
    {On leave of absence from Institute of Physics, Baku, Azerbaijan}
\\
\\Joint Institute for Nuclear Research,
\\ Dubna, Russia}          

\title { Light output simulation of LYSO single crystal}   

\begin{document}
\maketitle
{
\vskip 0.5cm
\hskip 6.5cm {\bf \large { Abstract}}
\vskip 0.5cm

{\large {
We used  the Geant4 simulation toolkit to estimate the light 
collection in a LYSO crystal by using   
cosmic muons and E=105 MeV electrons.
The light output as a function of the crystal length is studied.
Significant influence of the crystal wrapping 
in the reflective paper and optical grease  coupling to the              
photodetectors on the light output is 
demonstrated.
\Large {
\section{Introduction}
\vskip -0.8cm
$ $

The role of the experiments searching for the lepton flavor violation to
constrain the models of new physics has increased  over           
two years of the LHC operation: any indications of new physics are still beyond the grasp of
experiments. Flavor changing by all neutral current interactions is strongly 
suppressed in the Standard Model\,(SM). The new physics scenarios beyond the SM 
(supersymmetry, extra dimensions, little Higgs, quark compositeness) naturally        
allow and predict the charged lepton flavor violation at some level (see, e.g.,\cite{nphy}).

The aim of the $\mu\to e$ conversion experiments is to search for the coherent
conversion of the muons from muonic atoms 
to the electrons in the field of a nucleus through some new 
lepton flavor violation interactions.
The conceptual designs\cite{come,mu2e} of the $\mu\to e$ conversion 
experiments include  the calorimeter  able
to measure the energy of the electrons with the resolution <5\% for 105\,MeV 
and the time resolution  $\sim$1\,ns to provide the
trigger signal and measure track positions in addition to the tracking chambers.
The calorimeter will consist of the 3x3\,cm$^2$   dense
crystals and are >10 radiation lengths long.
Several factors affect the accuracy of the energy and time measurements using
the scintillator-light detector.
The energy resolution  of a photopeak at energy E
can be expressed in terms of  various contributions\cite{birk1,tawa}
$$ {R_{tot}^2 } = {R_{ph}^2 + R_{sc}^2 + R_{en}^2},$$
where $R_{tot}$ is the full width at half maximum energy resolution,
$R_{ph}$ represents the contribution from the photon   statistics, $R_{sc}$
represents the influence from the nonideal nature of the crystal (inhomogeneities,
nonproportionality), variance in the light-collection efficiency, and
$R_{en}$ represents the contribution of electronic noise.                      
The nature of the reflective
wrapping around the crystal and the interface between
the crystal and the light detector must also be considered as  the
factors affecting the quality of measurements: the former leads to the increase    
and the later to the loss  of the scintillation photons collection.
For an ideal scintillator and ideal readout electronics  $R_{sc}=R_{en}=0$ and
for $N_{ph}$ photoelectrons 

$$R_{tot}  =  R_{ph}=2.35/{\sqrt{N_{ph}}.}$$                                     
The quantity $N_{ph}$ can be expressed in  terms of
the mean number of scintillation photons, $N_{\gamma}$, produced by the scintillator due to
the absorption of energy, 
the collection efficiency of the detector system for scintillation photons,
${\epsilon}_{\gamma}$, the  collection efficiency for the photoelectrons, ${\epsilon}_{e}$,   
the  quantum efficiency of the associated photon detectors, $q_{eff}$,
and the total gain of the light detector, $g_t,$ 

$$N_{ph} = N_{\gamma}{\epsilon}_{\gamma}{\epsilon}_{e} q_{eff}g_{t}.$$
The analysis of  different contributions
to the energy resolution for some crystals was performed, e.g., in\cite{marc,herb}.
\begin{figure}[ht!]                            
{\vskip -1.0cm}
{\hskip  1.6cm}  {\epsfxsize  5.0 truein \epsfbox{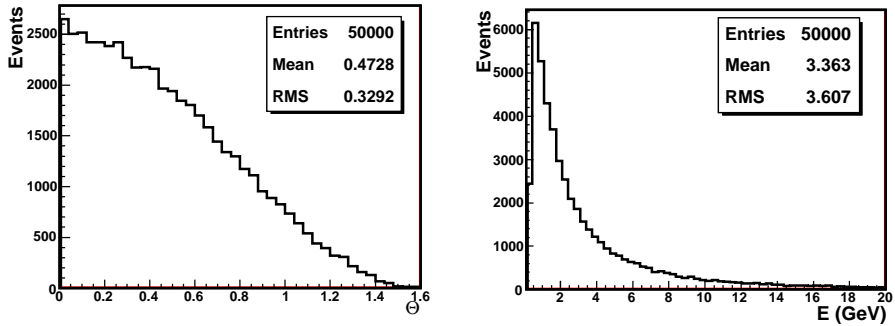}}
{\vskip -2.2cm}
\caption[]{ The azimutal angular distribution (left) and energy spectrum (right) of cosmic muons as
simulated according to\cite{volko}.}                                                                        
\label{Norm}
\end{figure}

The reliable Monte-Carlo simulation plays an important role in the
crystal selection and the calorimeter configuration, minimizing the systematic
uncertainties in the physics analysis. 
Without knowing   certain detector properties and neglecting
the effects of crystal inhomogeneity,  we restricted this work 
to the  simulation of the 
number of photons  in the LYSO crystal.     
The next section gives a brief description of the LYSO crystal choice advantages.
Section~3 describes the cosmic muon simulation  and the strategy
of simulating  optical photons in a crystal using  Geant4 toolkit.
In Section~4 we give the light output results. The dependence of the 
photon number on the size of the crystal and influence of the crystal wrapping is studied.
In this section we also show the energy deposition in the crystal when cosmic
muons and E=105 MeV electrons are used.
We end with the conclusions in Section~5.

\section{Inorganic crystal for the calorimeter}
\vskip -0.8cm
$ $

Recently\cite{usu1} we have performed the comparative analysis of three dense crystals
for using in the  $\mu \to e$ conversion experiment.
Several superior characteristics ---
high stopping power( radiation length $X_0$=1.14\,cm, $R_{Moliere}$=2.07\,cm), 
fast decay time($\tau$=40\,ns),  bright 
scintillation(light yield $\ge$26\,photons/keV)
relative to many other crystals used in high energy physics,
nonhygroscopic nature, and radiation hardness  --- make cerium-doped
lutetium and lutetium-yttrium oxyorthosilicats (LSO:Ce and LYSO:Ce) 
major candidates for being used  in trigger calorimeters in  the $\mu \to e$-conversion
experiments\footnote{The unpleasant factors are the natural radioactivity and 
 a  high price of LSO/LYSO crystals
in relation to commonly used scintillators,
which may be significantly reduced if they are mass produced
for high-energy physics needs.}.

The commonly used optical photon detectors  have 
high $q_{eff}$ at the peak emission wavelength, 420\,nm,
of LSO/LYSO  crystals. For example, the $e_{eff}$ values                            
for the  Photonis XP2254B     
photomultiplier  tube and the Hamamatsu S8664 avalanche photodiode      
are $7.2\pm 0.4\%$ and $75\pm 4\%$\cite{mao1}, respectively.      
Luminescence properties  and compatibility of the LYSO:Ce crystal 
to many currently employed optical photon detectors 
are discussed in more  detail in\cite{vala1}.

\begin{figure}[ht!]                            
{\vskip -0.0cm}
{\hskip  2.1cm}  {\epsfxsize  5.0 truein \epsfbox{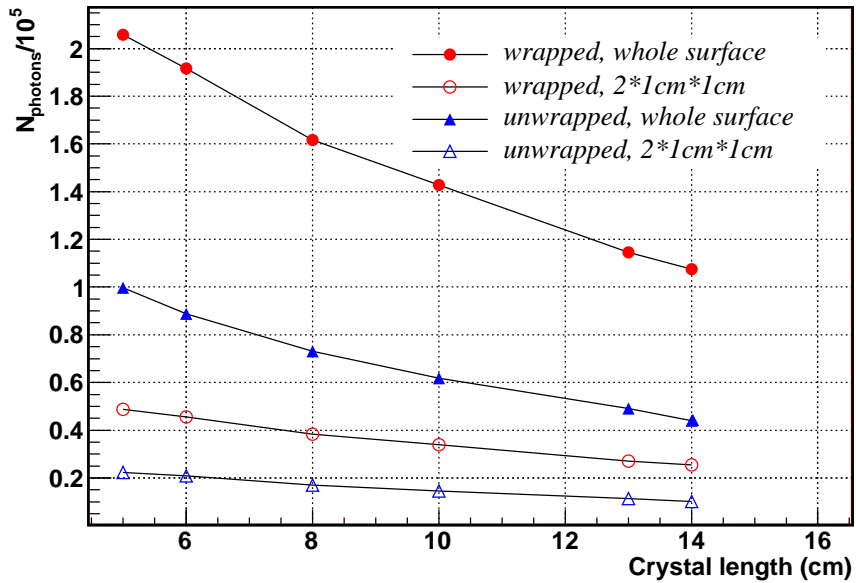}}
{\vskip -0.0cm}
\caption[]{The number of photons collected from the whole 3x3 cm$^2$ surface and two 1x1 cm$^2$
areas as a function of the LYSO:Ce crystal length (see the text).
Cosmic muons penetrate the crystal  
perpendicular to the 3x13 cm$^2$ surface at a random point.}
\label{Norm}
\end{figure}

\begin{figure}[ht!]                            
{\vskip -0.0cm}
{\hskip  0.5cm}  {\epsfxsize  6.0 truein \epsfbox{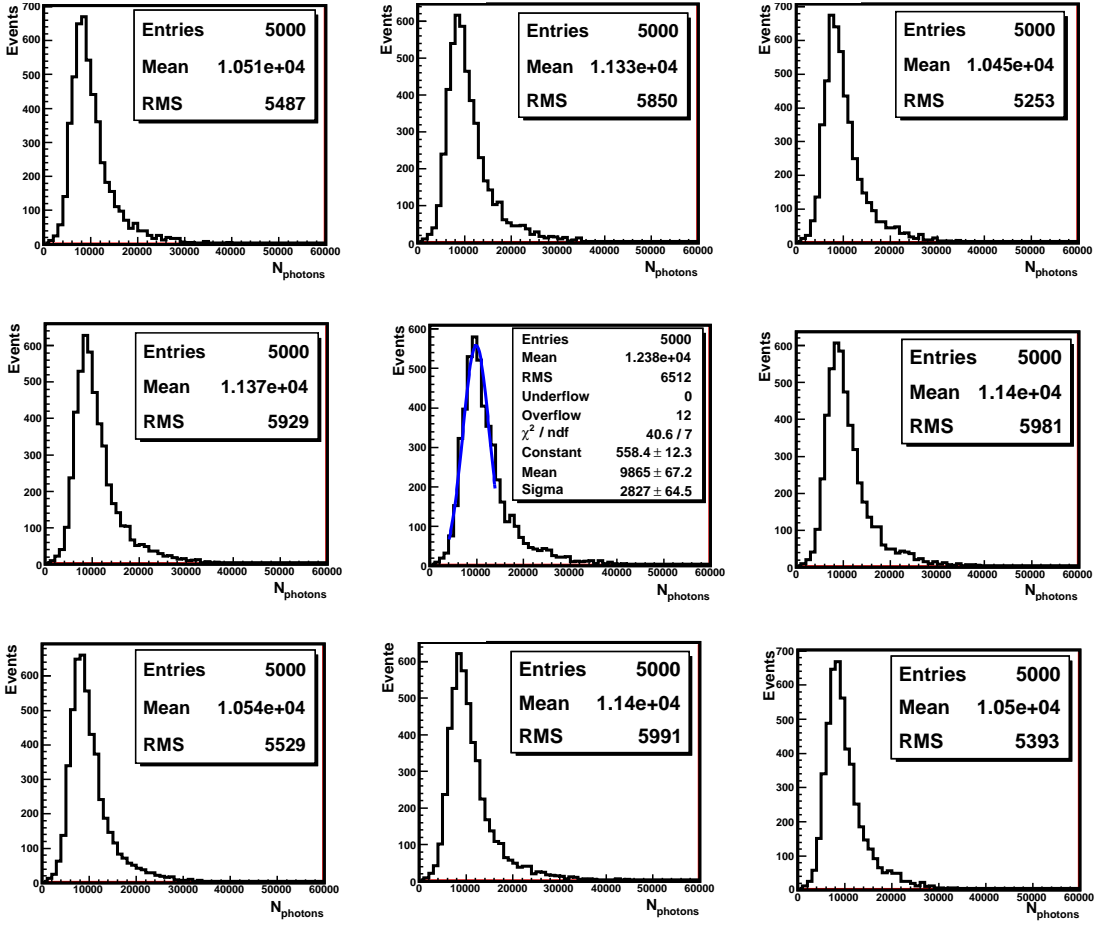}}
{\vskip -0.0cm}
\caption[]{The number of photons in the nine areas into which the 
3x3 cm$^2$ crystal  surface is divided.
The E=105 MeV electrons penetrate the center of the crystal on the opposite side
in the perpendicular direction.}
\label{Norm}
\end{figure}

\begin{figure}[ht!]                            
{\vskip -0.0cm}
{\hskip  2.1cm}  {\epsfxsize  5.0 truein \epsfbox{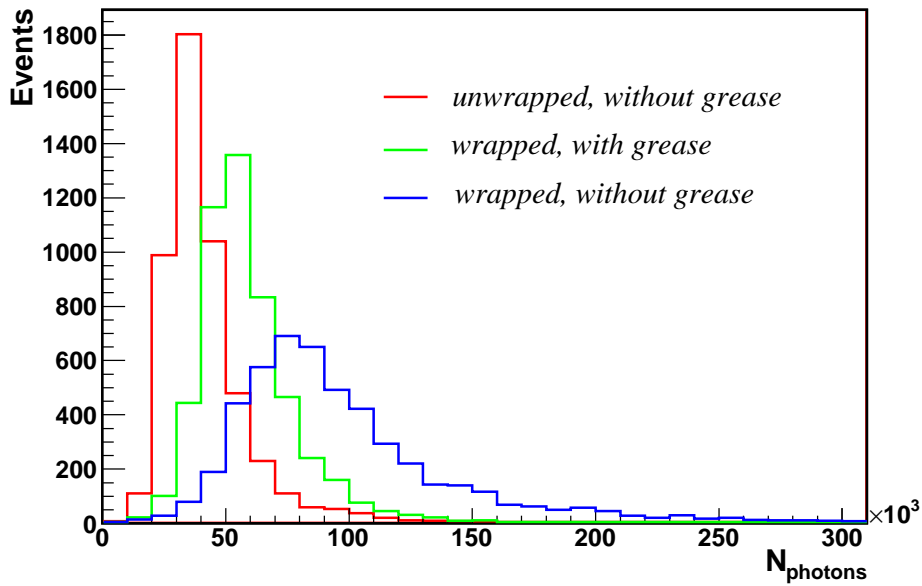}}
{\vskip -0.0cm}
\caption[]{The number of photons collected from the whole 3x3 cm$^2$ surface when 105 MeV electrons
penetrate the center of the opposite surface of the crystal in the perpendicular directions.
The influence of wrapping in Tyvek paper and optical grease are demonstrated.}                             
\label{Norm}
\end{figure}

\section{Cosmic muons and optical photons simulations}
\vskip -0.8cm
$ $

The accurate simulation of the processes in the scintillator 
was performed using optical
and scintillation models of  Geant4\cite{geant} version 9.6.p01. The Geant4
low-energy electromagnetic libraries 
were  employed in that simulation.
The studied crystal is LYSO:Ce 
with the dimensions 3x3x13 cm$^3$. 
The yttrium content of LYSO:Ce is 4\%, the cerium doping level is 0.02\%. The refractive
index  of LYSO:Ce was set as a function of the optical photon wavelength\cite{mao2}.
We assumed the intrinsic light output of the LYSO:Ce crystal to be 26 scintillation                     
photons per keV (see, e.g.,\cite{laan}). The photon absorption length in whole interval  
of wavelengths was put to  20\,cm\cite{vilar,kripl}.

The crystal is polished on all surfaces. From one of the readout ends (3x3\,cm$^2$)
we collected the photons. The opposite surface of the crystal is covered by nonreflective
black paper. Four lateral layers of the crystal were 
wrapped in highly reflective (R=97\%)               
Tyvek paper to collect  photons effectively. 
Reflection of  photons
from the surfaces between two dielectric materials was simulated  using
the UNIFIED
model\cite{geant}. We used the polished and ground types of the surfaces.
We also performed simulation without wrapping  the crystal.

   The minimal tracking step in the simulation was set to 
10$\,\mu$m, which corresponded to the energy
cuts of $\sim$3.2~keV for the photons  
and $\sim$51.9(50.8)~keV for the electrons(positrons)
for the LYSO:Ce  crystals.

The cosmic muons were generated according to the energy spectrum\cite{volko} 
and injected perpendicularly to the 3x13 cm$^2$
surface of the crystal at a random point. 
The azimuthal angular distributions and energy spectrum
of the simulated cosmic  muons are shown in Fig.\,1.         
In the range 0.3-1000\,GeV 
 <E$_{cosm}$>=9.79\,GeV.

\section{Results}                                          
\vskip -0.8cm
$ $

In this study the light from the single LYSO:Ce crystal is collected 
(1) from the whole 3x3 cm$^2$  surface       
divided into nine 1x1 cm$^2$ areas
and (2) from two 1x1 cm$^2$ areas arranged symmetrically about the  vertical 
axis of the crystal  at a distance of 0.5 cm from each other. The lateral surfaces of      
the crystal was covered  with  Tyvek paper without a thin air gap.                              

Figure\,2 compares the number of photons collected from whole 
crystal surface and two 1x1 cm$^2$ areas
for the naked and wrapped LYSO:Ce crystal with  different lengths.                                            
The cosmic muons were injected perpendicularly to the upper 3x13 cm$^2$ side 
of the crystal at a random point. Note that
Tyvek wrapping increases the light collection by a factor of more than 2.  The number of
photons decreases by a factor of $\sim$1.4 when the length of the crystal increases from 5 to 10 cm.
We have found  that the use  of the previous Geant4 version (9.5.p01) leads to a decrease  in the 
number of photons by 3.7\% and 2.2\%
for the crystal lengths 5 and 13 cm, respectively.

In Fig.\,3   we present the number of photons collected from the whole 3x3 cm$^2$ 
surface of the LYSO:Ce crystal 
divided into  nine 1x1 cm$^2$ areas. The E=105\,MeV electrons  are 
injected into the center of the opposite 3x3 cm$^2$ surface.
The maximal number of photons  is  collected in the central part of the crystal, 
and the minimal number was collected in its corners.
The same behavior is observed if the cosmic muons are injected perpendicularly to the  
upper surface of the crystal at a random point. The ratio $\sigma$/E 
for the central cell corresponding to the Gaussian fit is $\sim$0.29\,(see\,Fig.\,3).

In Fig.\,4 we show the number of photons collected from  the 
13-cm-long naked and wrapped LYSO:Ce crystals  using   105\,MeV               
electrons. The electrons were injected perpendicularly into the center of the 3x3 cm$^2$ surface 
of the crystal and the photons were  collected from the whole opposite surface of the crystal.
The influence of  0.1\,mm optical grease is also demonstrated in the figure.
The optical grease
(polidimethilsiloxane, C${_2}$H${_6}$OSi,
\,$\rho$=0.97\,g/cm$^3$,\,reflective index  R=1.4) is used for attaching the 
photodetectors to the crystal  and transmitting  light to the photodetector.
From the Gaussian fit, the mean of photon numbers
corresponding
to the wrapped crystal, wrapped crystal with optical grease, and naked crystal 
are in the ratio    
1:0.57:0.42. Note that between the reflective paper and the crystal there is  no thin
air gap and without grease the light is collected just after crystal.

Finally in Fig.~5 we demonstrate the energy deposition in the LYSO:Ce crystal by  cosmic
muons and  E=105 MeV electrons. The cosmic muons and electrons penetrate the crystal
as described above. The Gaussian fit curves and parameters are also demonstrated in the figure.
The ratio $\sigma / E $ for these distributions is  0.091   and 0.061 
for the cosmic muons and electrons,   respectively.
\begin{figure}[ht!]                            
{\vskip -0.0cm}
{\hskip  2.1cm}  {\epsfxsize  5.0 truein \epsfbox{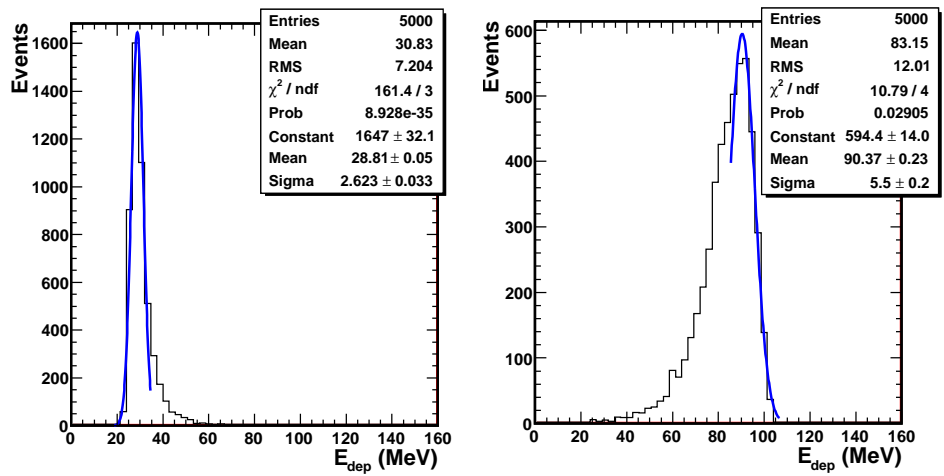}}
{\vskip -0.0cm}
\caption[]{The energy deposition in the LYSO:Ce crystal when cosmic muons penetrate the 3x13 cm$^2$ surface
in the perpendicular direction at a random point (left)  and when 105 MeV electrons penetrate
the center of the 3x3 cm$^2$ surface (right).}
\label{Norm}
\end{figure}

\section{Conclusions}                                            
\vskip -0.8cm
$ $

In this paper we focused our attention on the light output of LYSO:Ce crystals 
neglecting their nonproportionality and inhomogeneity and light detector response.
The simulation of the crystal is based on the Geant4 simulation toolkit. 
The light output was studied
using  cosmic muons and E=105 MeV electrons. 
Significant influence of the size and  wrapping 
in  reflective paper on the light output is 
demonstrated. The optical grease used for attaching the photodetectors to the crystal
leads to a decrease   in  the light output.
The central region of the crystal on the end surface is the
most predominant area for photon collection.

To validate the simulation, these results should be benchmarked with
experimental measurements of LYSO:Ce crystals.
Such comparisons should lead to the correct considerations of nonproportionality
of the scintillator response  below 300\,keV and crystal inhomogeneity in the simulation.
The experimental data on the light spectra and photon numbers for 
LSO:Ce and LYSO:Ce crystals obtained 
with  different light detectors also are necessary for correct simulation.

}
\

}}

\begin{thebibliography}{99}
\bibitem{nphy}
J.\,Hisano and D.\,Nomura, Phys. Rev. D59, 116005 (1999);  \\
K.\,Agashe, A.E.\,Blechman and F.\,Petriello, Phys. Rev. D74, 053011 (2006);\\
M.\,Blanke et al., JHEP 0705, 013 (2007).
\bibitem{come} 
K.\,Akhmetshin et al., Letter of Intend for Phase-I of the COMET Experiment
at J-PARC, KEK/J-PARC-PAC 2011-27, March 11, 2012.
\bibitem{mu2e}
R.M.\,Carey et al.,"Proposal to Search for $\mu^{-}N \to e^{-}N$ with 
a Single Event Sensitivity Below $10^{-16}$", FERMILAB-PROPOSAL-0973.
\bibitem{birk1}
J.B.\,Birks, The Theory and Practice of Scintillation Counting, Pergamon Press, London 1964.
\bibitem{tawa}
H.\,Tawara, S.\,Sasaki, K.\,Saito, and E.\,Shibamura,
Proceeding of the Second International Workshop on EGS, August 2000, Tsukuba, Japan
KEK Proceeding 200-20, p.152-160.
\bibitem{marc}
C.W.E.\,van\,Eijk, P.\,Dorenbos, E.V.D.\,van\,Loef, K.\,Kr\"amer, and H.U.\,G\"udel,
Radiat. Meas., 33, 521 (2001).
\bibitem{herb}
D.J.\,Herbert, L.J.\,Meng, and D.\,Ramsden, IEEE Transactions on Nuclear Science, 49, 931 (2002).
\bibitem{usu1}
Z.\,Usubov, Electromagnetic calorimeter simulation for future $\mu \to e$ conversion
experiments, arXiv:1212.4322[physics.ins-det].
\bibitem{mao1}
R.\,Mao, L.\,Zhang, and R.-Y.\,Zhu, IEEE Transactions on Nuclear Science, 55, 1759 (2008).
\bibitem{vala1}
I.G.\,Valais et al., IEEE Transactions on Nuclear Science 54(1), 11 (2007).                         
\bibitem{geant}
S.\,Agostinelli et al., Nucl.Instr. and Meth. A506, 250 (2003); \\
GEANT4 Collaboration, Geant4 User's Guide for Application Developers, Dec. 2010.
\bibitem{mao2}
R.\,Mao, L.\,Zhang and R.-Y.\,Zhu, IEEE Transactions on Nuclear Science 55, 2425 (2008).
\bibitem{laan}
D.J. van der Laan et al., Phys. Med. Biol. 55, 1659 (2010).
\bibitem{vilar}
I.\,Vilardi et al., Nucl. Instr. and Meth. A564, 506 (2006).
\bibitem{kripl}
A.\,Kriplani et al., Conference Record Proceedings, paper M14-183, 2003
IEEE Nuclear Science Symposium and Medical Imaging Conference, 
Oct. 19-26, Portland, OR, (2003).
\bibitem{volko}
L.N.\,Volkova, G.T.\,Zatsepin, L.A.\,Kuzmichev, Yad. Fiz. 29, 1252 (1979).
\end{thebibliography}
\end{document}